\documentstyle[11pt,fleqn]{article}
\begin{document}

\begin{titlepage}

\begin{flushright}
Freiburg--THEP 96/6\\
April 1996
\end{flushright}
\vspace{1.5cm}

\begin{center}
\large\bf
{\LARGE\bf On the evaluation of three--loop
           scalar integrals in the massive case}\\[1cm]
\rm
{Adrian Ghinculov}\\[.5cm]

{\em Albert--Ludwigs--Universit\"{a}t Freiburg,
           Fakult\"{a}t f\"{u}r Physik}\\
      {\em Hermann--Herder Str.3, D-79104 Freiburg, Germany}\\[1.5cm]
      
\end{center}
\normalsize

\begin{abstract}

I describe a method to calculate a class of three--loop selfenergy diagrams
for arbitrary internal masses and external momentum. This method combines
analytical results and numerical integration, and is suitable for
implementation in a computer program to produce fast and 
accurate results. For the class of Feynman diagrams considered 
in this paper this method leads to a two--fold integral 
which needs to be evaluated numerically. Results are
given for a range of masses and external momentum.
\end{abstract}

\vspace{3cm}

\end{titlepage}


\section{Introduction}

Perturbation theory based on Feynman diagrams continues to be the main tool
for calculating observable quantities which can be measured in 
high energy experiments. Unfortunately, the difficulty of the calculations
increases dramatically with the loop order.

Some electroweak parameters are being measured in LEP/SLC experiments 
at the per mille level. To match such accuracy, full two--loop 
electroweak corrections may be needed in the future \cite{preccalc}.
Higher loop contributions are also needed when one deals with
relatively large couplings. 
Here higher order corrections can be used to  increase the accuracy 
of the calculation and to estimate the validity range of perturbation theory.
In the electroweak sector potentially large radiative corrections may
be induced for instance by the large Yukawa coupling of the top quark,
and also if the mass of the minimal Higgs boson turns out to be large.
One can also imagine extensions of the symmetry breaking sector of
the standard model where strong interactions may play an essential r\^{o}le
\cite{singletmodel}.

In calculating radiative corrections, the difficult step is usually
the evaluation of the scalar integrals. 

At one--loop order this
is in principle a completely solved problem. However, in practice the
evaluation of one--loop scalar integrals in the general mass case can
be tricky because one has to make sure the functions involved remain
on the physical Riemann sheet, and because of the need to control
potentially large numerical cancellations \cite{vermold}.

At two--loop level, well--known results by van der Bij and Veltman
guarantee that all vacuum diagrams with arbitrary masses 
can be expressed analytically 
in terms of Spence functions \cite{vdBij:2loop:rho}. 
If the external momenta 
do not vanish, this is in general not possible any longer.
In fact, the two--loop selfenergy diagram with three propagators
is known to be related to the Lauricella function, which is
a generalization of the hypergeometric series \cite{lauricella}. 
Such functions 
are not straightforward to evaluate. Typically they  
are expressed by multiple series whose convergence 
may be poor for certain choices of the parameters, 
for instance near a threshold. Therefore one has to rely 
at the two--loop level at least partly on numerical techniques.
A combination of analytical and numerical methods exists which can 
be used in principle to calculate any two--loop diagram. 
There exist also a large number of techniques which work
for certain topologies or for special choices of parameters.
On the evaluation of two-loop scalar integrals in the general
massive case see refs. \cite{lauricella}---\cite{fujimoto2}.

While techniques to handle massless three--loop diagrams
exist \cite{mincer}, the massive case is much more difficult.
Apart from a formal expression for the four propagator selfenergy
diagram \cite{lauricella}, no general results exist at the three--loop level.
Considering the previous remarks on the massive two--loop 
diagrams, there is little hope that analytical techniques can
prove useful for realistic three--loop calculations of physical relevance
in the massive case.

It is the purpose of this letter to provide efficient 
methods for evaluating 
certain massive three--loop diagrams. 
Instead of trying to solve the general problem, this paper limits itself
to calculating the selfenergy diagram of fig. 1. 
With little modifications, this method can be used to
calculate some other diagrams, of the type shown in fig. 2.

The scalar integral corresponding to the diagram of fig.1 is:

\begin{eqnarray}
\lefteqn{I(m_1^2,m_2^2, \dots ,m_8^2 ; k^2) \, =}         \nonumber \\
   &  & \int d^{n}p\,d^{n}q\,d^{n}r\, 
        \frac{1}{(q^2 - m_1^2) [(q+k)^2 - m_2^2]}\,
        \frac{1}{(r^2 - m_4^2) [(r-k)^2 - m_5^2]}
       \nonumber \\
  & &   \frac{1}{(p^2 - m_7^2) [(p+k)^2 - m_8^2]}\,
        \frac{1}{[(p-q)^2 - m_3^2] [(p+r)^2 - m_6^2]}
    \; \; .
\end{eqnarray}

To calculate this diagram one starts by performing the integrals 
over $q$ and $r$ analytically, and leaves the $p$ integral to be
performed numerically after an examination of the analytical
structure of the integrand.

The $q$ and $r$ integrals are essentially two one--loop triangle
subdiagrams. Of course, in
our case two of the external momenta of the triangle subdiagrams 
are functions of the loop momentum $p$ of the three--loop diagram.
To fix the notations, let us denote:

\begin{eqnarray}
\lefteqn{C(m_1^2,m_2^2,m_3^2; p_1^2,p_2^2,p_3^2) \, =}         \nonumber \\
     & & \int d^{n}p\, 
        \frac{1}{(p^2 - m_1^2) [(p-p_2)^2 - m_3^2] [(p+p_3)^2 - m_2^2]}
    \; \; .
\end{eqnarray}

Such triangle graphs are expressible in terms of Spence functions for 
any choice of the internal masses and external momenta. Explicit
formulae can be found for instance 
in refs. \cite{veltthooft} or \cite{denner}. These expressions
were encoded in the program FF \cite{FF}. This program will be used
in the following to calculate the diagram of fig. 1 in the general mass case.
However, the rather lengthy expression of the function $C$ simplifies 
considerably for special choices of the masses. Such is for instance 
the massles case, $m_1=m_2=m_3=0$. In this case a formula can
be derived which is better suited for numerical evaluation than
the general case formula used in FF. 
Equivalent formulae were obtained previously by other authors
(see for instance ref. \cite{davydychev:vertex}), but the expression
which will be derived in the following has the advantage of remaining
on the right sheet for the parameter range needed in this
three--loop calculation. 
We therefore sketch the derivation of
this formula. 

One introduces Feynman parameters, performs the integral over the loop
momentum, expands in $\epsilon = n-4$, integrates over one Feynman
parameter, and obtains:

\begin{equation}
C(0,0,0; p_1^2,p_2^2,p_3^2) =  - i \frac{\pi^2}{p_3^2}  \int_{0}^{1} dx \,
  \left[ \,  \frac{1}{x (1-x) (1-\mu^{2})} \log \mu^{2} \, \right]
      \; \; ,
\end{equation}
where

\begin{eqnarray}
   \mu^{2}  & = &  \frac{a x + b (1-x)}{x (1-x)}   \nonumber \\
         a  & = &  \frac{p_{1}^{2}}{p_{3}^{2}} \, , \; \; \; \;
         b \; = \; \frac{p_{2}^{2}}{p_{3}^{2}} \, , \; \; \; \;
      \; \; .
\end{eqnarray}

It suffices to keep only the ${\cal O}(\epsilon^0)$ terms in the expression of
$C$ because the three--loop diagram of eq. 1 is obviously 
ultraviolet convergent. The integral of eq. 3 can be carried out analytically
and gives:

\begin{eqnarray}
\lefteqn{C(0,0,0; p_1^2,p_2^2,p_3^2) \, =}         \nonumber \\
& &    \frac{2 \, i \, \pi^2}{p_3^2 \, \sqrt{\Delta}} \,
     \left[ \,  Sp(-\frac{u_{2}}{v_{1}}) + Sp(-\frac{v_{2}}{u_{1}}) 
     + \frac{1}{4} \log^{2} \frac{u_{2}}{v_{1}}
     + \frac{1}{4} \log^{2} \frac{v_{2}}{u_{1}} \right.
                                                \nonumber \\  
& &   \; \; \; \; \; \; \; \; \; \; \; \; \left.
      + \frac{1}{4} \log^{2} \frac{u_{1}}{v_{1}}
      - \frac{1}{4} \log^{2} \frac{u_{2}}{v_{2}}
      + \frac{\pi^{2}}{6} 
   \,  \right]
      \; \; ,
\end{eqnarray}
where

\begin{eqnarray}
u_{1,2}        & = & \frac{1}{2}
              ( 1 + b - a
                    \pm \sqrt{\Delta} )  \nonumber \\
v_{1,2}        & = & \frac{1}{2}
              ( 1 - b + a
                    \pm \sqrt{\Delta} )  \nonumber \\
\Delta  & = & 1 - 2 (a+b) + (a-b)^{2}
      \; \; .
\end{eqnarray}

This function is similar to the finite part of the two--loop integral
\linebreak
${\cal G}(m_1,m_2,m_3;0)$ of ref. \cite{2loop}, or $(m_1,m_1|m_2|m_3)$ 
of ref. \cite{vdBij:2loop:rho}. 
This is not surprising, since there is a relation
between one--loop massless diagrams evaluated at finite external momentum
and massive two--loop vacuum diagrams. Other relations of this type
were obtained for instance in ref. \cite{davydychev:vertex}.

One can now write the three--loop diagram of fig. 1 in the following form:

\begin{eqnarray}
\lefteqn{I(m_1^2,m_2^2, \dots ,m_8^2 ; k^2) \, =}         \nonumber \\
     &  & \int d^{4}p \, 
        \frac{1}{(P_1^2 - m_7^2) [P_2^2 - m_8^2]}\,
        C(m_1^2,m_2^2,m_3^2;P_2^2,P_1^2,k^2)\, \nonumber \\
     &  & \;\;\;\;\;\;\;\;\;\;\;\;\;\;\;\;\;\;
	 \times C(m_4^2,m_5^2,m_6^2;P_2^2,P_1^2,k^2)
    \; \; ,
\end{eqnarray}
where

\begin{eqnarray}
P_1^2  & = &  p^2  
\nonumber \\ 
P_2^2  & = & (p+k)^2
      \; \; .
\end{eqnarray}

This expression has the structure of a one--loop integral with 
some vertex form factors. As already mentioned, one does not need to keep
terms of order $\epsilon$ and higher in the expression of $C$ because
$I$ is ultraviolet finite. One can choose the external momentum to have
vanishing space components, $k=(\sqrt{k^2},\vec{0})$. 
Let us denote the time and space components of the loop momentum
by $p \equiv (p_0,\vec{p})$. It is obvious that the integrand in eq. 7
is independent of the direction of $\vec{p}$. Therefore one can readily 
perform the angular integration over the direction of $\vec{p}$. 
The result reads:

\begin{eqnarray}
\lefteqn{I(m_1^2,m_2^2, \dots ,m_8^2 ; k^2) \, =}         \nonumber \\
     &  & \int_{-\infty}^{\infty} d p_0\,\int_{0}^{\infty} d \rho\, 
        \frac{4 \, \pi \, \rho^2}{(P_1^2 - m_7^2) [P_2^2 - m_8^2]}\,
        C(m_1^2,m_2^2,m_3^2;P_2^2,P_1^2,k^2)\, \nonumber \\
     &  & \;\;\;\;\;\;\;\;\;\;\;\;\;\;\;\;\;\;
	 \times C(m_4^2,m_5^2,m_6^2;P_2^2,P_1^2,k^2)
    \; \; ,
\end{eqnarray}
where

\begin{eqnarray}
P_1^2  & = &  p_0^2 - \rho^2  
\nonumber \\ 
P_2^2  & = & p_0^2 + 2 p_0 \sqrt{k^2} + k^2 - \rho^2
      \; \; ,
\end{eqnarray}
and $\rho = |\vec{p}|$.

In this way one has obtained a two--fold integral representation of the
three--loop diagram shown in fig. 1.

Some remarks on the numerical integration are now in place.

The possible singularities of the integrand in the integration domain
$(p_0,\rho) \in (-\infty,+\infty) \times (0,+\infty)$
are given by the solutions of the equations:

\begin{eqnarray}
p_0^2 - \rho^2 - m_7^2 + i \eta & = &  0  
\nonumber \\ 
p_0^2 + 2 p_0 k + k^2 - \rho^2 - m_8^2  + i \eta & = &  0
      \; \; ,
\end{eqnarray}
and by the thresholds of the vertex functions $C$. The threshold
behaviour of the functions C does not pose special problems with
respect to the numerical integration of expression 9 because the integrand
remains finite at these points. One also has to keep in mind that
for special values of the mass and momentum parameters, additional problems 
may come from the infrared singularities of the function
$C$. $C(m_1^2,m_2^2,m_3^2;p_1^2,p_2^2,p_3^2)$ displays an 
infrared singularity for $p_3^2=m_2^2$, $p_2^2=m_3^2$ and $m_1^2=0$
and cyclic permutations. 

The simplest and most efficient way to avoid 
the singularities associated with eqns. 11
is to perform a Wick rotation of the loop momentum $p$. Instead of 
the usual rotation of the time component $p_0$, it is more convenient to rotate 
the radial space component $\rho$ with a {\em negative} phase $e^{-i\alpha}$,
with $\alpha \in (0,\pi/2)$. This way the integrand becomes free 
of singularities and can be integrated easily.
At the same time this procedure allows a useful check on the calculation
because the result must be independent of the actual value of $\alpha$.

One can also calculate this 
integral without introducing a complex $\rho$. To do this, the integral 
has to be split along the solutions of eqns. 11. Without this factorization
of the singularities, the adaptative integration algorithms are rather 
inefficient because the singularities are smeared across both integration 
variables $p_0$ and $\rho$.

As a check on the calculation, one notices that modifying formula 9
into:

\begin{eqnarray}
\lefteqn{J(m_1^2,m_2^2, \dots ,m_5^2 ; k^2) \, =}         \nonumber \\
     &  & \int_{-\infty}^{\infty} d p_0\,\int_{0}^{\infty} d \rho\, 
        \frac{4 \, \pi \, \rho^2}{(P_1^2 - m_4^2) [P_2^2 - m_5^2]}\,
        C(m_1^2,m_2^2,m_3^2;P_2^2,P_1^2,k^2)
    \; \; ,
\end{eqnarray}
gives the two--loop selfenergy diagram shown in fig. 3, which was
calculated by a number of authors. For $m_1=m_2=m_3=0$, $m_4=m_5=|k|=1$
the result is:

\begin{eqnarray}
\lefteqn{J(0,0,0,1,1;1) \, =}         \nonumber \\
   & &  \pi^4 \, [  \, (1.8472631 \pm 1.8 \cdot 10^{-6})
                + i \, (3.4451413 \pm 1.8 \cdot 10^{-6}) \, ]
    \; \; ,
\end{eqnarray}
in agreement with already existing results \cite{2loop,kniehl}. 
The case
$m_1=m_2=m_3=m_4=m_5=|k|=1$ gives:

\begin{equation}
J(1,1,1,1,1;1)  =  \pi^4 \, [  \, (.923663 \pm 2.6 \cdot 10^{-5})  \, ]
    \; \; ,
\end{equation}
which again agrees with the known value of this diagram \cite{2loop,kniehl}.

Turning now to the three--loop case, fig. 4 shows  the results for the
selfenergy diagram of fig. 1 for $m_1=m_2=m_3=m_4=m_5=m_6=0$, $m_7=m_8=1$
as a function of the external momentum $k^2$. 
It was checked that both eq. 5 and the general formula encoded
in FF lead to the same results.
Fig. 5 shows the same diagram in the all--massive case
$m_1=m_2=m_3=m_4=m_5=m_6=1/2$, $m_7=m_8=1$.

Obviously, it is straightforward to modify the formula 9 for calculating 
other three--loop diagrams of the type shown in fig. 2. In a similar way,
it is also possible to introduce more propagators in eq. 12 for 
calculating a few other two--loop scalar integrals.
For example,
fig. 6 shows the results for diagram 2 a)  
in the case $m_1=m_2=m_3=m_4=m_5=m_6=0$, 
$m_7=1$. 

To conclude, a simple method was described which allows one to calculate
efficiently a class of three--loop selfenergy diagrams. It works for 
any masses and momentum combinations in the physical region. 
As a by--product,
this yields a new method for calculating certain two--loop Feynman diagrams.


\vspace{.5cm}

{\bf Acknowledgement}

I am grateful to the theory department of Brookhaven National
Laboratory, where this paper was written during a short visit, 
for hospitality. 
This research was supported by the Deutsche Forschungsgemeinschaft (DFG).


\newpage


{\bf Figure captions }

\vspace{2cm}

{\em Fig.1}    Three--loop selfenergy diagram in the general massive case. 

\vspace{.5cm}

{\em Fig.2}    Some three--loop diagrams which can be calculated with the
               same methods as the diagram in fig. 1. The dots on internal
	       lines are external vertices connected to zero external
	       momenta.

\vspace{.5cm}

{\em Fig.3}    A two--loop selfenergy diagram which can be calculated
               with the same methods as the three--loop diagram of fig. 1.

\vspace{.5cm}

{\em Fig.4}    The real (solid line) and the imaginary (dashed line) parts
               of the scalar three--loop function $\pi^6 \, I$,
	       with $I$ defined in eq. 1, as a function
	       of the external momentum squared $k^2$. The masses of the 
	       internal lines are $m_1=m_2=m_3=m_4=m_5=m_6=0$, $m_7=m_8=1$.
	       Note that due to the definition of $I$ the absorbtive part
	       of the corresponding diagram is the {\em real} part of $I$.

\vspace{.5cm}

{\em Fig.5}    Same as fig. 4, but for the all--massive case
               $m_1=m_2=m_3=m_4=m_5=m_6=1/2$, $m_7=m_8=1$.

\vspace{.5cm}

{\em Fig.6}    Same as fig. 4, but for the three--loop diagram 
               of fig. 2 a) in the case case
               $m_1=m_2=m_3=m_4=m_5=m_6=0$, $m_7=1$.

\end{document}